\documentclass[11pt,a4paper]{article}
\pdfoutput=1
\usepackage{jcappub}

\usepackage[inline,margin,marginclue,index]{fixme}
\makeatletter
\renewcommand*\FXLayoutInline[3]{%
  {\@fxuseface{inline} \ignorespaces[#3 \fxnotename{#1}: #2]}}
\makeatother
\fxsetup{theme=color,mode=multiuser} \FXRegisterAuthor{Javi}{ame}{\color{blue}Javi}
\fxsetup{theme=color,mode=multiuser} \FXRegisterAuthor{todo}{am}{\color{green}todo}
\fxsetup{theme=color,mode=multiuser} \FXRegisterAuthor{Igor}{a}{\color{red}Igor}
\fxsetup{theme=color,mode=multiuser} \FXRegisterAuthor{Babette}{b}{\color{brown}Babette}

\usepackage{bbold}
\usepackage{tikz}
\usepackage{epsfig}
\usepackage{amssymb}
\usepackage{color}
\usepackage{framed}

\renewcommand\({\left(}
\renewcommand\){\right)}

\newcommand{\be}{\begin{equation}}
\newcommand{\ee}{\end{equation}}
\newcommand{\bea}{\begin{eqnarray}}
\newcommand{\eea}{\end{eqnarray}}
\newcommand{\exclude}[1]{}

\newcommand{\vvvv}[4]{ \left( \begin{array}{cccc}   #1 \\ #2 \\ #3 \\ #4\end{array} \right)}
\newcommand{\vvvvv}[5]{ \left( \begin{array}{cccc}   #1 \\ #2 \\ #3 \\ #4 \\ #5\end{array} \right)}

\newcommand{\gagamma}{g_{\scriptscriptstyle A\gamma}}

	\newcommand{\qomega}{\Omega}
	\newcommand{\fomega}{\omega}
	\newcommand{\qomegaC}{{\tilde \qomega}}
	\newcommand{\fomegaC}{{\tilde \fomega}}
	\newcommand{\Emod}{{\cal E}\hspace{-6pt}{\cal E}}
	\newcommand{\kk}{K}
	\newcommand{\E}{\Xi} 
	\newcommand{\cev}{\overline}
	\newcommand{\geo}{{\cal G}}
	\newcommand{\MM}{\mathbb{M}}
	\newcommand{\eig}{e}
	\newcommand{\OO}{\mathbb{\Omega}}
	\newcommand{\jA}{J_A}
	\newcommand{\ii}{j}

\begin{document}


\subheader{CERN-OPEN-2018-001, MPP-2018-18}

\title{Axion Searches with Microwave Filters: the RADES project}

\author[a]{Alejandro \'Alvarez Melc\'on,}
\author[b]{Sergio Arguedas Cuendis,}
\author[c]{Cristian Cogollos,}
\author[a]{Alejandro D\'iaz-Morcillo,}
\author[b]{Babette D\"obrich,}
\author[d]{Juan Daniel Gallego,}
\author[e]{Benito Gimeno,}
\author[c]{Igor G. Irastorza,}
\author[a]{Antonio Jos\'e Lozano-Guerrero,}
\author[b]{Chlo\'e Malbrunot,}
\author[a]{Pablo Navarro,}
\author[f,g]{Carlos Pe\~na Garay,}
\author[c,h]{Javier Redondo,}
\author[b]{Theodoros Vafeiadis,}
\author[b]{Walter Wuensch.}

\affiliation[a]{Department of Information and Communication Technologies, Universidad Politecnica de Cartagena, Murcia, Spain}
\affiliation[b]{European Organization for Nuclear Research (CERN), 1211 Geneva 23, Switzerland}
\affiliation[c]{Departamento de F\'isica Te\'orica, Universidad de Zaragoza, 50009, Zaragoza, Spain}
\affiliation[d]{Yebes Observatory, National Centre for Radioastronomical Technologies and Geospace Applications, Guadalajara 19080, Spain}
\affiliation[e]{Department of Applied Physics and Electromagnetism-ICMUV, University of Valencia, Spain}
\affiliation[f]{I2SysBio, CSIC-UVEG, P.O.  22085, Valencia, 46071, Spain}
\affiliation[g]{Laboratorio Subterr\'aneo de Canfranc, Estaci\'on de Canfranc, 22880, Spain}
\affiliation[h]{Max-Planck-Institut f\"ur Physik (Werner-Heisenberg-Institut), 80805 M\"unchen, Germany}

\abstract{We propose, design and construct a variant of the conventional axion haloscope concept that could be competitive
in the search for dark matter axions of masses in the decade 10-100 $\mu$eV.
{Theses masses are located somewhat above the mass range in which existing experiments have
reached sensitivity to benchmark QCD axion models.}
{Our haloscope} consists of an array of small microwave cavities connected by
rectangular irises, in an arrangement commonly used in radio-frequency filters. The size of the unit cavity determines the main
resonant frequency, while the possibility to connect a {large} number of cavities allows to reach large detection volumes.
We develop the theoretical framework of the detection concept, and present design
prescriptions to optimize detection capabilities. {We describe the design and realization of a first small-scale prototype of this concept, called Relic Axion Detector Exploratory Setup (RADES). It consists of a copper-coated stainless steel five-cavities microwave filter with the detecting mode operating at around 8.4 GHz.}
This structure has been electromagnetically characterized at 2~K and 298~K, and it is now placed in ultra-high vacuum in one of the twin-bores of the 9~T CAST dipole magnet at CERN. We describe the data acquisition system
developed for relic axion detection, and present preliminary results of the electromagnetic properties of the microwave
filter, which show the potential of filters to reach QCD axion window sensitivity at X-band frequencies.
 }

\maketitle
\flushbottom

\listoffixmes

\section{Introduction}

Axions, as well as more generic axion-like particles (ALPs), are currently considered one of the
most promising portals for new physics beyond the Standard Model (SM) of particle physics.
Axions arise in extensions of the SM including the Peccei-Quinn (PQ) mechanism \cite{Peccei:1977hh,Peccei:1977ur}, currently the most
compelling solution \cite{Weinberg:1977ma,Wilczek:1977pj} to the strong-CP problem of Quantum Chromo Dynamics (QCD). More generic ALPs often appear in diverse types of SM extensions. Not necessarily related to the axion,
ALPs share part of its phenomenology. For example, it is now known that string theory naturally predicts many ALPs (and the axion itself) \cite{Cicoli:2012sz}.
Beyond their motivation from theoretical arguments, there are additional arguments motivating their existence coming from cosmology and astrophysics.
Most relevantly, axions are strong candidates to compose all or part of the dark matter (DM).

Indeed, non-relativistic axions
could have been produced in the early
universe by the phenomenon called \textit{vacuum-realignment} (VR) and, in addition, by the decay of topological defects (TD) of the axion field, like domain walls and axion strings \cite{Kawasaki:2014sqa}. For both mechanisms, the production is approximately inversely proportional to the axion mass, which means that the condition of overproduction of relic axion density translates into a lower bound on the axion mass. However, the computation of the relic axion density for a given axion model and, correspondingly, the axion mass for which the right DM density is obtained, is in general rather uncertain. This is due to dependencies on axion cosmology model and, in the case of the TD, on computational difficulties.

For axion models with PQ transition happening before inflation, only the VR contribution needs to be considered (TD are removed by inflation),
but it turns out to be dependent on the unknown value of the initial misalignment angle $\theta_i$, unique for all the observable Universe.
Assuming a natural $\mathcal{O}$(1) value for this angle, it would translate
to axion masses at the $\sim10~\mu$eV scale, but much lower mass values could be justified by anthropically finetuned values of $\theta_i$~\cite{Tegmark:2005dy}.
For axion models with PQ transition happening after inflation, the VR contribution is calculated over an averaged value of $\theta_i$, thus removing
the previous uncertainty. Therefore for these models one can safely set a lower bound to the axion mass of, at least, $m_A \gtrsim 25 \mu$eV, but possibly higher, depending on the
importance of the TD contribution. However this contribution is difficult to compute, as it requires detailed numerical simulation of the behavior of the defects.
A recent study \cite{Kawasaki:2014sqa} claims that TD contribution is dominant, and therefore shifts the right axion mass up to the 80--130 $\mu$eV.  A more recent attempt to quantify this contribution \cite{Klaer:2017ond} provides a lower range $m_A = 26.2  \pm 3.4~\mu$eV.  In addition, one must note that in particular models in which the TD are long-lived, due to the existence of several almost-degenerate vacua in the axion potential, the TD contribution can be substantially increased, and therefore the right axion mass can go to much larger values \cite{Kawasaki:2014sqa}. Moreover, if the axion is a subdominant part of DM, the axion mass also moves to correspondingly higher values. On more general grounds, the VR mechanism is common also to generic ALPs, and a large fraction of the ALP parameter space can also potentially contain viable ALP DM models~\cite{Arias:2012az}. To summarize, although a large mass-range is in principle open to axion DM exploration, there is a specific motivation to extend the sensitivity of conventional searches, so far competitive in the low mass range 1-10 $\mu$eV, to higher values.

The conventional axion haloscope technique~\cite{Sikivie:1983ip} consists of a high-$Q$ microwave cavity inside a magnetic field to trigger the conversion of axions from our 	galactic DM halo into photons.
Being non-relativistic, the axions convert to monochromatic photons with energy equal to $m_A$. For a cavity whose resonant frequency matches $m_A$, the conversion is enhanced by a factor proportional to the quality factor of the cavity $Q$.
For a high $Q$ cavity,
{the resonant band is small and thus}
the cavity must  be tunable and data taking is performed by scanning very thin mass-slices of parameter space. Therefore, a useful figure of merit $\mathcal{F}$ of these
experiments is proportional to the time needed to scan a fixed axion mass range~\cite{Asztalos:2001tf} down to a given signal-to-noise level and for a given value of the axion-photo coupling $\gagamma$:

\begin{equation}
\label{FOM}
    \mathcal{F} \sim \gagamma^4 m_A^2 B^4 V^2 T_{sys}^{-2} \geo^4 Q
\end{equation}

\noindent where $B$ is the magnetic field (assumed constant over the cavity volume), $V$ is the cavity volume, $T_{sys}$ is the detection noise temperature,  and $\geo$ is the
geometrical form factor of the cavity mode, typically proportional to the overlap integral between the mode electric field and the external magnetic field.

The ADMX
collaboration \cite{Asztalos:2009yp} has demonstrated that this technique is competitive in the 1 to 10 $\mu$eV range and it has realistic prospects to explore this
range down to the QCD axion sensitivity in the near future.
Pushing these prospects to higher masses is challenging, because it requires to make the cavity resonant to higher frequencies, which means a reduced volume $V$, and correspondingly reduced sensitivity. One can in principle compensate the loss in $V$ by improving other parameters, like $Q$, $B$ or $T$ and indeed substantial effort in these directions is ongoing in the community. But a most appealing option would be to effectively increase $V$ by filling a large volume with many high-frequency resonant structures, i.e. effectively decoupling the detection volume $V$ from the volume of a cavity and the resonant frequency. Literally replicating a cavity many times and combining their output is possible in theory, although difficult in practice due to the need to phase match them,
and probably it will be challenging to scale it above a few cavities. An alternative approach is to design extended periodic structures that could in principle fill large volumes while coherently resonating at a high frequency. The need of tuning the resonant frequency and keeping competitive values for the rest of parameters still makes this option challenging,
although promising recent ideas are being tested \cite{TheMADMAXWorkingGroup:2016hpc,Rybka:2014cya,Miceli:2015xas}.

We here propose and develop another particular realization of this idea, in which the high-frequency resonant structure is an array of $N$ small rectangular cavities connected with irises. Such an arrangement (see Figure \ref{fig:cavity}) resembles that of
a radio-frequency (RF) filter, although, as will be seen later, it differs in its design parameters. We find that this concept allows for an (in principle) arbitrarily large magnetic volume to be instrumented with cavities, while the resonant frequency is (mostly) determined by the size of a single element. One single readout channel is foreseen for the full array, i.e avoiding the need of challenging offline signal combination and phase matching. In addition, the geometrical layout of such arrangement could be realized in different ways, providing flexibility in instrumenting a large magnetic volume.
The motivation for considering this setup is manifold.
First, we would like to develop a technique that prioritizes scaling-up in $V$, even at the cost of
trading-off in other parameters, like, e.g. $Q$. The goal is to take advantage of large magnetic volumes in already existing infrastructures that could be made available to this type of research.
The CAST magnet is one such existing example. It has been already used for $\sim$15 years for solar axion searches and is now partially devoted to host axion haloscope test setups. It is a 10-meter long,
2$\times$15~cm$^2$ aperture, 9 T superconducting dipole, which corresponds to a total $B^2 V = 2.4$~T$^2$m$^3$ (to be compared, e.g., to the $B^2 V \sim 11$~T$^2$m$^3$ of the ADMX magnet).
The ambitious goal would be to instrument something like the magnet of the future axion helioscope IAXO \cite{Armengaud:2014gea}, a dedicated toroidal magnet with $B^2 V \gtrsim 300$~T$^2$m$^3$ with similarly large aspect ratio.
These numbers are very promising to extract axion DM signals provided ways to efficiently use this magnetic volume are developed.
Second, arranging the cavities in 1-dimensional arrays is perfectly suitable to instrument
magnets with large aspect-ratios like the ones of CERN accelerator magnets or the future IAXO.
Third, the weak linear coupling between cavities through small irises allows a very simple theoretical description that can greatly aid the design of multiple cavities.
{Fourth, tuning haloscopes composed of multiple subcavities can be really challenging. To simplify the design we considered many cavities optimized around a central frequency with simple and robust tuning mechanisms to allow retuning in a small range around the central peak ($\sim10\%$). This strategy would be compromised if the design phase was long and complicated; the theoretical guidance described in the following sections is a key part of the conceptual design adopted.}

The concept here proposed is being experimentally tested as the Relic Axion Detector Exploratory Setup (RADES) project. RADES is part of the new experimental program of CAST, presented and approved
by CERN SPSC in 2015 \cite{spsc2015}, and now under implementation. As a first step, a small-scale RADES cavity with 5 elements and no tuning has been built and installed inside the CAST magnet
for operation in the current data taking campaign of the experiment. We must note that another complementary idea to make use of large aspect-ratio magnetic volumes,
based on the use of single long rectangular cavities \cite{Baker:2011na}, is also being tested at the
CAST magnet by the CAST-CAPP team.
{Furthermore, the cavity array concept independently proposed in \cite{Goryachev:2017wpw} shares some conceptual elements with our proposal. In particular, the authors of \cite{Goryachev:2017wpw} identify some of the merits of cavity structures that will be developed in our work, in particular regarding the scalability in $V$ for high frequency operation. We here further develop the concept by providing a prescription to optimize the coupling of the structure with the axion field. Moreover, we propose a different practical implementation based on a filter-like structure (while in \cite{Goryachev:2017wpw} a series of posts or coils are proposed instead), and we present feasibility results, both based on complete simulations and on a first experimental prototype. Finally, we recently realised that a very similar implementation to the one here discussed, an array of cavities interconnected with irises, was already mentioned by D. Morris in an unpublished preprint~\cite{Morris:1984nu} from 1984. }


In section \ref{sec:modelling} we will present the theoretical background of the concept of a set of
individual cavities with a coupling between neighboring ones. We will develop design prescriptions to optimize the performance
of the array as axion detectors. In section \ref{sec:design} we discuss possible implementations of this concept based on full
3D simulations of the cavity array. In section \ref{sec:RADESdesign} we present the first RADES demonstrative prototype, currently
operated in the CAST magnet at CERN, and give sensitivity expectations in section \ref{sec:sensi}.
We finish in section \ref{sec:summary} with our conclusions and discussion of prospects.

\section{Theoretical modeling of a microwave filter \label{sec:modelling} }

In this paper we refer to an array of cavities connected by irises as a filter. Furthermore, for simplicity, we
consider only the excitation of the fundamental mode of each subcavity, i.e. higher harmonics are assumed to be
well separated. The excitation of a filter by the oscillating axion DM field\footnote{Recall that the frequency of the axion DM field is similar to the axion mass $\omega\sim m_A$, which is a priori unknown. We work in natural units $\hbar=c=1$. }, $A = A_0 e^{-\ii \omega t}$, $\ii$ being the imaginary unit, can be described as
\begin{align}
\label{theequationt}
(\omega^2 {\mathbb{1}} - \MM)\cev{\E} = \cev {\jA} = - \gagamma B_e A_0\,  \omega^2 \, \cev \geo,
\end{align}
where $\cev \E$ is a column vector of the $E$-field amplitudes of the fundamental mode of each cavity, $\cev {\jA}$ is the vector whose components are the axion DM excitation parameters of each cavity $(\cev {\jA})_q=- \gagamma B_e A_0 \, \omega^2 \geo_q$, with $\geo_q$ being the geometric factor of the cavity defined in \eqref{geo}($\cev \geo$ a vector of all of them).
We derive the formula \eqref{theequationt} in appendix~\ref{sec:genform}, where we detail also a little more on the theoretical aspects of coupled cavities.
The matrix $\MM$ contains the  natural frequencies, damping factors and couplings between cavities.
In our case of rectangular cavities segmented and connected through irises, $\MM$ is modeled by the tri-diagonal and symmetric matrix
\begin{equation}
\label{OMt}
\MM =
\left(
\begin{array}{c c c c c c}
\qomegaC_{1}^{2} & \kk_{12} & 0 & 0 & 0 & 0 \\
\kk_{21} & \qomegaC_{2}^{2} & \kk_{23} & 0 & 0 & 0\\
0 & \kk_{32} & \qomegaC_{3}^{2} & \kk_{34} & 0 & 0 \\
0 & 0 & \ddots & \ddots & \ddots & 0 \\
0 & 0 & 0 & \ddots & \ddots & \ddots \\
0 & 0 & 0 & 0 & \kk_{N,N-1} & \qomegaC_{N}^{2} \\
\end{array}
\right)  ,
\end{equation}

Neglecting losses (which will always be kept very small), $\MM$ is a real square symmetric matrix of dimensions $N \times N$, with $N$ real eigenvalues with associated eigenvectors.
The eigenvalues correspond to the square of the $N$ resonant eigenfrequencies, $\fomega_i^2$, and the eigenvalues
are vectors $\cev \eig_i$ representing the $E$-field amplitude and phase of the fundamental mode of each of the individual cavities.
In our notation, $\qomega_q$ is the eigenfrequency of the $q$-th \emph{individual} cavity in the limit of $\kk \to 0$,
and $\fomega_i$ stands for the $i$-th resonant frequency of the filter as a whole.
Note that we use the label $q$ for properties of the individual cavities and $i$ for those of the global filter.
A tilde above $\fomega,\qomega$ denotes a complex frequency, where the imaginary part accounts for losses.
Also, $\kk_{q-1,q}$ parametrizes the coupling between the $q-1$ and $q$ cavities.
The solution of \eqref{theequationt} gives the electric fields in each cavity $\cev{\E}$ as a superposition of the $E$-fields of the resonant modes,
\bea
(\cev \E)_q
		&=& \sum_{i} (\cev \eig_i)_q\(\frac{\cev \eig_i \cdot \cev {\jA}}{\omega^2-\fomegaC_i^2}\) \,
\eea
see the appendix for the derivation.

\begin{figure}[h!]
\begin{center}
\includegraphics[height= 6cm,width = 0.9\textwidth]{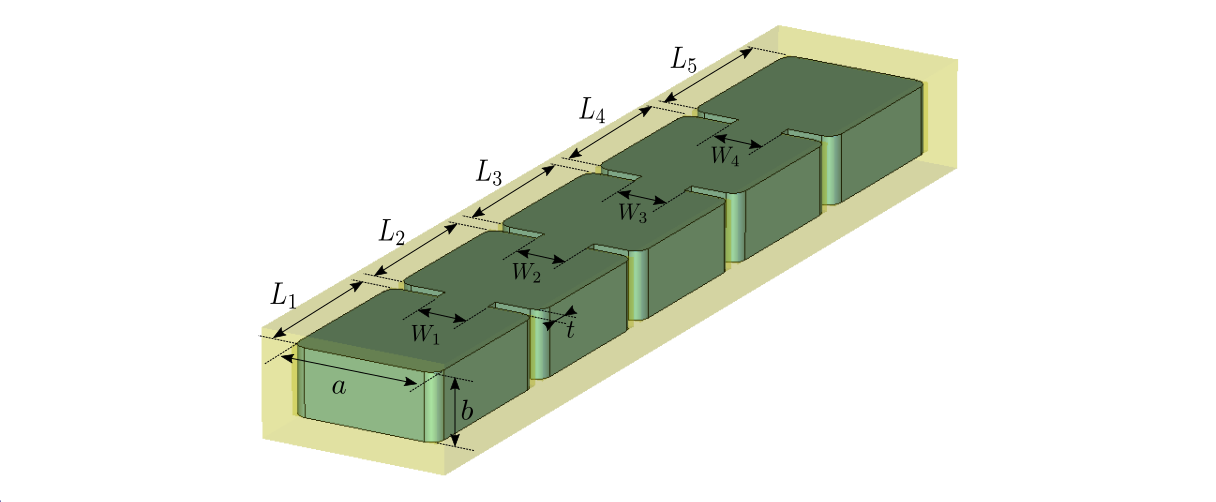}
\caption{{Design of a RADES microwave filter-like structure} for axion detection composed by five coupled cavities with length $L_i$ and dimensions $a$ and $b$, connected by rectangular irises of width $W_i$ and depth $t$.}
\label{fig:cavity}
\end{center}
\end{figure}

For practical implementation, we have selected a filter with 5 cavities and 4 couplings, which we
show in Figure~\ref{fig:cavity}.
The eigenvalue problem associated with this matrix for a given eigenvalue $\fomega_i^2$, leads
to the following system of linear equations
\begin{equation}
\label{EIG}
\fomega_i^2
\left(
\begin{array}{c}
\eig_{i1}\\
\eig_{i2}\\
\eig_{i3}\\
\eig_{i4}\\
\eig_{i5}\\
\end{array}
\right)
 =
 \MM
 \left(
 \begin{array}{c}
 \eig_{i1}\\
 \eig_{i2}\\
 \eig_{i3}\\
 \eig_{i4}\\
 \eig_{i5}\\
 \end{array}
 \right)
 =
  \left(
 \begin{array}{c}
			  \qomega_{1}^{2}\,\eig_{i1} +\kk_{12}\,\eig_{i2}\\
\kk_{12}\,\eig_{i1}+\qomega_2^2\,\eig_{i2}+\kk_{23}\,\eig_{i3}\\
\kk_{23}\,\eig_{i2}+\qomega_{3}^2\,\eig_{i3}+\kk_{34}\,\eig_{i4}\\
\kk_{34}\,\eig_{i3}+\qomega_{4}^2\,\eig_{i4}+\kk_{45}\,\eig_{i5}\\
\kk_{45}\,\eig_{i4}+\qomega_{5}^2\,\eig_{i5}\\
 \end{array}
 \right)
\end{equation}
where we use $\eig_{iq}=(\cev \eig_{i})_q$ ($i,q=1, 2 \cdots 5$) for the components of the corresponding
eigenvector. By introducing  $k^{(i)}_{q-1,q} = {\kk_{q-1,q}}/{\fomega_i^2}$ the
following explicit system of linear equations is obtained
%
\bea
\label{EIGV} \nonumber
   \fomega_i^2 &=& \frac{\qomega_{1}^{2}\,\eig_{i1} }{\eig_{i1}-k^{(i)}_{12}\, \eig_{i2}}
   = \frac{\qomega_{2}^{2}\,\eig_{i2} }{\eig_{i2}-k^{(i)}_{12}\, \eig_{i1}-k^{(i)}_{23}\, \eig_{i3}}
   = \frac{\qomega_{3}^{2}\,\eig_{i3} }{\eig_{i3}-k^{(i)}_{23}\, \eig_{i2}-k^{(i)}_{34}\, \eig_{i4}}\\
   && = \frac{\qomega_{4}^{2}\,\eig_{i4} }{\eig_{i4}-k^{(i)}_{34}\, \eig_{i3}-k^{(i)}_{45}\, \eig_{i5}}
   = \frac{\qomega_{5}^{2}\,\eig_{i5} }{\eig_{i5}-k^{(i)}_{45}\, \eig_{i4}}.
\eea

We emphasize that we can choose the design parameters $\qomega_q$'s and $k_{q-1,q}$'s by
altering the dimensions of the cavities and irises.
The idea thus, is to find out for which values of the parameters we can obtain a filter coupling
to the axion DM with a given set of characteristics.
{For this first} work we have chosen to fix a desired characteristic frequency $\fomega_i$ (the operational frequency),
and design a filter that maximizes the geometric factor for that frequency,
\be
\label{geot}
\geo_i =
\frac{\sum_q  V_q\,  \cev \eig_i \cdot \cev \geo}{V} ,
\ee
see discussion around \eqref{geoN} in the appendix.
In this equation we can observe that the form factor depends on the alignment of the electric field in each cavity to the external
magnetic field, ${\bf B}_e$, which is here represented by the sign of $\geo_q$.
Consequently, it can be maximized by forcing the electric field in all five cavities to be aligned
with the applied magnetic field.
The geometric factors of each individual cavity in a constant external magnetic field $B_e$
are very similar. This is because they are similar in size and working on the fundamental
mode so the vector of geometric factors is $\cev \geo \simeq  (1,1,1,1,1) \times \geo$, with
$\geo$ the geometric factor of a cavity. Maximum coupling can thus be obtained when one of
the filter eigenmodes is $\cev e = (1,1,1,1,1)/\sqrt{N}$.

There is some freedom to realize this, but the simplest solution that we have found is to take all coupling coefficients to be equal to a fixed value $k^{(i)}$ with $k^{(i)}=k^{(i)}_{12}=k^{(i)}_{23} = k^{(i)}_{34} =k^{(i)}_{45}$.
We choose $k^{(i)}<0$, for which our desired solution is the lowest-frequency mode, min$\{\fomega_i\}=\fomega_1$.
For positive $k^{(i)}$, our desired mode would correspond to $\fomega_5$.
The resonant frequencies of the individual cavities $\qomega_q$ can now be computed from Eq.~(\ref{EIGV}), obtaining
\begin{equation}
  \qomega_2^2 = \qomega_3^2 = \qomega_4^2=\fomega_1^2 \,(1 - 2 k^{(1)}) \quad , \quad
            \qomega_1^2 =  \qomega_5^2 =  \fomega_1^2 \,(1 - k^{(1)}) \,.
\label{design_eq}
\end{equation}
The solution is extremely simple: all cavities must share the same resonant frequency, except for the first and the last one,
which need to have larger resonant frequencies by a factor $\qomega_1/\qomega_2=\sqrt{(1-k^{(1)})/(1-2k^{(1)})}$, determined by the selected coupling value.
For negative $k^{(1)}$, the factor is smaller than one, so the external cavities have to be slightly larger than the internal ones.
Actually, this solution holds for arbitrary $N$!

It is straightforward to compute the 4-remaining modes. However, before doing so we can already advance that they do not couple to the axion, i.e. $\cev \eig_{2,3,4,5}\cdot \geo = 0$. The reason is very easy to understand. In absence of losses, the eigenvectors form an orthogonal basis of the vector space of $E$-fields in the cavities. Since we have already chosen one vector to lie parallel to the excitation vector $\cev \geo \propto (1,1,1,1,1)$ the rest are orthogonal to it, and thus uncoupled.
Analytically, we obtain
\begin{align}
\frac{\fomega^2_i}{\fomega^2_1} &= 1, \quad\quad 1 - \frac{3-\sqrt{5}}{2}k^{(1)}, \quad 1 - \frac{5-\sqrt{5}}{2}k^{(1)}, \quad
1 - \frac{3+\sqrt{5}}{2}k^{(1)}, \quad 1 - \frac{5+\sqrt{5}}{2}k^{(1)} \\
\label{eigenvectorsRADES}
\cev \eig_1 &\propto \vvvvv{1}{1}{1}{1}{1} , \,
\cev \eig_2 \propto \vvvvv{+1}{+\varphi-1}{0}{-\varphi+1}{-1}, \,
\cev \eig_3 \propto \vvvvv{-1}{2 - \varphi}{2\varphi-2}{2 - \varphi}{-1}, \,
\cev \eig_4 \propto \vvvvv{-1}{+\varphi}{0}{-\varphi}{1}, \,
\cev \eig_5 \propto \vvvvv{-1}{+\varphi+1}{-2\varphi}{+\varphi+1}{-1}
\end{align}
where $\varphi=(1+\sqrt{5})/2=1.61803$ is the golden ratio.
Note that at this level of approximation the eigenvectors do not depend on $k^{(1)}$.

The figure of merit introduced in Eq.~(\ref{FOM}) also depends on the quality factor
of the filter. The \emph{unloaded} quality factor, $Q_i^u$, is defined as the ratio of the stored
EM energy in a mode, $U_i$, to the intrinsic power losses (due to finite conductivity of the cavity walls) per cycle, $P^c_i$~\cite{Collin:2001},
\begin{equation}\label{QF}
Q_i^u=   \frac{U_{i}}{P^c_i/\omega_i} = \frac{\omega_i}{\Gamma_i^c}
\end{equation}
where $\Gamma_i = - {\rm Im}\{\fomegaC^2_i\}/\fomega_i$ is the power decay rate of the mode, and the superscript denotes the
intrinsic losses of the cavity.  For a TE$_{101}$, one finds\footnote{Note that in \cite{Baker:2011na} a factor of 2 is missing in front of the $w^3$ term in formula (2.14) (arXiv), which appears as (2.15) in the PRD version. }~\cite{Baker:2011na}
\be
\label{unloaded101}
Q^u_q = \frac{1}{\delta} \frac{a b L(L^2+a^2)}{L a (L^2+a^2)+2 b(L^3+a^3)} ,
\ee
with $\delta$ the skin depth of the cavity walls, and the rest of notation referring to the cavity geometry (see Fig.~\ref{fig:cavity}).
The performance assessment of the proposed solution shown in the next section includes the estimation of this
parameter. In the appendix we show that
\be
\Gamma_i \simeq  \sum \Gamma_q \eig^2_{iq},
\ee
where $\Gamma_q=-{\rm Im}\{\qomegaC^2_q\}/\qomega_q$ is the power decay rate of the mode in cavity $q$.
If all the $\Gamma_q's$ were exactly the same, all filter modes would have $\Gamma_i=\Gamma_0$ and thus
$Q^u_i=Q^u_q$
because of orthonormality of the basis $\{\cev \eig_i\}$. However, the 1st and last cavity have only one iris, and therefore
more losses so some difference is expected.

\section{Design of a microwave filter for axion detection \label{sec:design}}
In this section we describe the design of five cavities microwave filter where the fundamental TE$_{101}$ mode is resonant in each cavity and optimized for axion detection operating at {a temperature of $\sim 2$ K, using the guidance developed} in previous section. {We start by fixing the desired frequency of operation of the system, i.e. that of the fundamental mode $\fomega_1$, and the inter-cavity coupling $k^{(1)}$. Our optimization condition in Eq.~\ref{design_eq} fixes the remaining parameters of the system.
We then need to translate the matrix elements $\qomega$'s and $K$'s of our analytical model of~\eqref{OMt} into physical dimensions, cavity and iris dimensions respectively}.

{We arbitrarily fix our frequency of operation to $\fomega_1=8.4$~GHz, as it corresponds to waveguide dimensions that comfortably fit into the CAST magnet bore. We restrict ourselves to a } WR90 EIA standard rectangular waveguide, which fixes the width $a$ and height $b$ of all our cavities as given in Table~\ref{tab:dimensions}.
{Within these conditions, the natural frequency $\qomega_q$ of cavity $q$ is determined by its length $L_q$:}

\be
\qomega_q^2 = \(\frac{\pi}{a}\)^2 + \(\frac{\pi}{L_q}\)^2.
\ee

{Note that this relation holds for an ideal isolated rectangular cavity. The presence of irises interconnecting the cavities, the presence of ports, or the fact that the corners are rounded (to facilitate machining) will introduce perturbations to the above relation. In general we need to resort to numerical simulation of the real geometry to precisely identify the value of $L_q$ corresponding to a given $\qomega_q$. This is done with} CST Microwave Studio electromagnetic commercial software package~\cite{cstmws}, which works with the time-domain Finite Integration Technique (FIT).

{A similar argument holds for the coupling $k^{(1)}$ and the iris dimensions.}
Each coupling coefficient $k^{(1)}$ can be identified with a set of irises dimensions: the width $W$ and length $t$ of the irises. We have fixed $t=2$ mm, due to mechanical constraints. The determination of the value of $W$ that corresponds to a given $k^{(1)}$ is achieved by numerical simulations. By standard calculations of inter-resonator coupling, using symmetry, two coupled resonators connected by an inductive iris can be divided into {two single resonators, one terminated by a magnetic wall and the other by an electric wall.}
The coupling $k^{(1)}$ is then determined from the resonant frequencies of the two individual resonators~\cite{Cameron:2007aa}. {Using this method, we can obtain $k^{(1)}$ for each physical width $W$}.
Finally, we must correct the lengths $L_q$ for the interaction between the cavities and the irises (loading effect), as reported in~\cite{Soto2010}.

{This procedure has been followed for a number of geometries exploring different values of $W$. The value chosen in Table~\ref{tab:dimensions} has been selected on grounds of practical convenience, i.e. good separation in frequency of the cavity modes and ease of construction. The above method gives a value of $k^{(1)}=-0.0185$ for the geometry chosen in Table~\ref{tab:dimensions}.}


\begin{table}[hbtp]
\begin{center}
\begin{tabular}{|c|c|c|} \hline
Dimensions [mm]          & $T=2$ K & $T=298$~K lengths \\
          & & (including the 30$\mu$m copper coating layer) \\ \hline \hline
 $a$                & 22.86 & 22.99 \\ \hline
 $b$                & 10.16 & 10.25 \\ \hline
 $L_1=L_5$          & 26.68 & 26.82 \\ \hline
 $L_2=L_3=L_4$          & 25.00 & 25.14 \\ \hline
 $W_1=W_2=W_3=W_4$  & 8.00 & 8.14 \\ \hline
 $t$                & 2.00 & 1.95 \\ \hline
\end{tabular}
\caption{Physical dimensions of the five cavities filter design at a temperature of 2~K and at room temperature, 298~K. In the latter case, dimensions include the 30~$\mu$m copper coating which was used in the construction of the RADES prototype, see section~\ref{sec:RADESdesign}.}
\label{tab:dimensions}
\end{center}
\end{table}

\begin{figure}[h!]
\begin{center}
\includegraphics[height= 8cm,width = 0.9\textwidth]{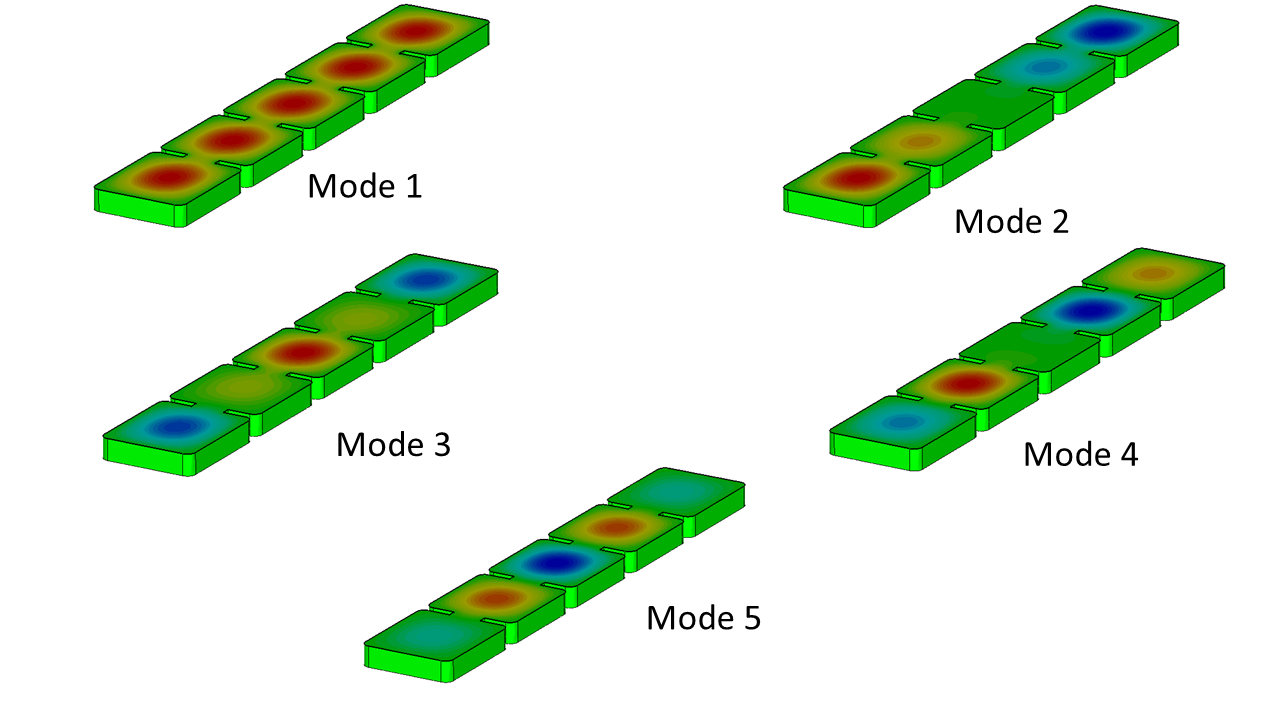}
\caption{Electric field distributions for the five characteristic
 modes of the designed filter. Observe that
 coherence between cavities is preserved only in the fundamental mode, as designed.}
\label{fig:efielddist}
\end{center}
\end{figure}
\begin{table}[hbtp]
\begin{center}
\begin{tabular}{|c|c|c|c|c|}\hline
Mode & Electric field pattern {(sign$(\cev \eig_i)_q$)}& $\fomega_i/2\pi$ (GHz)& $\geo^2_i$ & $Q^u_i$ (2~K) \\ \hline \hline
1& + + + + + & {8.428} & 0.65 & 40386\\ \hline
2& + + 0 - -& 8.454& 3.2$~10^{-7}$ & 42033\\ \hline
3& - + + + - & 8.528& 8.1$~10^{-5}$ & 43654\\ \hline
4& - + 0 - + & 8.625& 1.6$~10^{-12}$ & 45882\\ \hline
5& - + - + - & 8.710& 6.4$~10^{-6}$ & 48048 \\ \hline
\end{tabular}
\caption{Electric field pattern (signs of eigenvector coefficients), resonant frequency (eigenvalue), geometric form factor and quality factor for each characteristic mode of the designed filter-like cavity at 2~K, obtained from the CST simulations referred to in the text. }
\label{tab:modes}
\end{center}
\end{table}

%

{The remaining geometrical dimensions are fixed by our optimization prescription. In practice, this optimal geometry is finetuned by performing numerical computation of the electric field patterns of the resonant modes, and iterating over different values for $L_{1/5}$ (the length of the first and last cavities) until maximizing the numerically-computed geometric factor $\geo_1$ of the fundamental mode. This process completes all geometric parameters of the filter as shown in Table~\ref{tab:dimensions}}.

For this particular design, Table~\ref{tab:modes} shows the electric field patterns (signs of eigenvector coefficients) and resonant frequencies (eigenvalues) of the five characteristic modes as computed by CST eigenvalue solver.
The electric field patterns of the five resonant modes of the filter are shown in Fig.~\ref{fig:efielddist}.
{Note that they agree very nicely with the eigenvectors obtained analytically \eqref{eigenvectorsRADES}. } In particular, the first mode maintains the coherence along all the cavities of the structure by design.
Table ~\ref{tab:modes} also shows the geometric factor $\geo_i$ of each mode, obtained by numerical post-processing of the computed electric field values within the cavities.
As intended, the geometric factor is maximal for the first characteristic mode, and very close to the theoretical expectation of a TE$_{101}$ mode resonating in a single cavity $\geo_q^2=(8/\pi^2)^2=0.657$. In addition, the $\geo_i$ factor of all the other modes are very close to zero. This gives us confidence that we have indeed identified the correct geometry corresponding to the optimal configuration of our analytical model. Furthermore, the resonant frequency of the fundamental mode agrees well with the designed frequency $\fomega_1=8.4$ GHz.


{The unloaded $Q$ factor can also be computed with the above simulations, by introducing appropriate wall losses. The numbers shown in  Table~\ref{tab:modes} are obtained using as input a conductivity of $2\times10^9$~S/m. These values approximately agree with formula \eqref{unloaded101}, by which $Q^u_{101}\simeq 5.5 {\rm mm} /\delta$, assuming the skin depth of copper at 2~K to be $\delta\simeq 0.1\mu$m. Nevertheless, as will be discussed in the next section, there are other effects that are not well captured by the simulations and that will push experimental $Q$ to lower values.}

{To summarize, we have determined a concrete geometrical implementation of a set of five inter-connected cavities that correspond to the optimal solution from the analytical model presented in previous section. Detailed numerical simulation reproduces the features expected from the model solution, in terms of eigenvectors and eigenvalues. Future work will go in the direction of studying the scalability in $V$ of this solution, as well as its robustness against small variations of geometrical parameters (mechanical tolerances). In addition, work is ongoing to better understand the translation of the analytical model parameters into physical features of the filter. }


\section{A first exploratory setup \label{sec:RADESdesign}}
%
In this section, we describe how we have built and characterized our first five cavities X-band filter optimized for axion searches. The filter implementation of the design in a realistic prototype requires some additional considerations: materials selection, physical dimensions at room temperature, and coaxial probes insertions.

\begin{figure}[hbtp]
\begin{center}
\includegraphics[height= 0.22\textwidth]{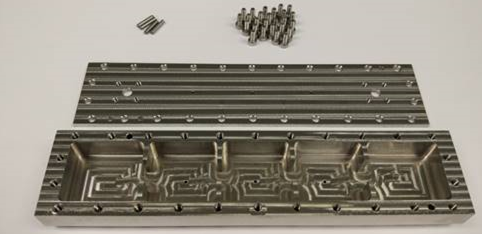}
\includegraphics[height =0.22\textwidth]{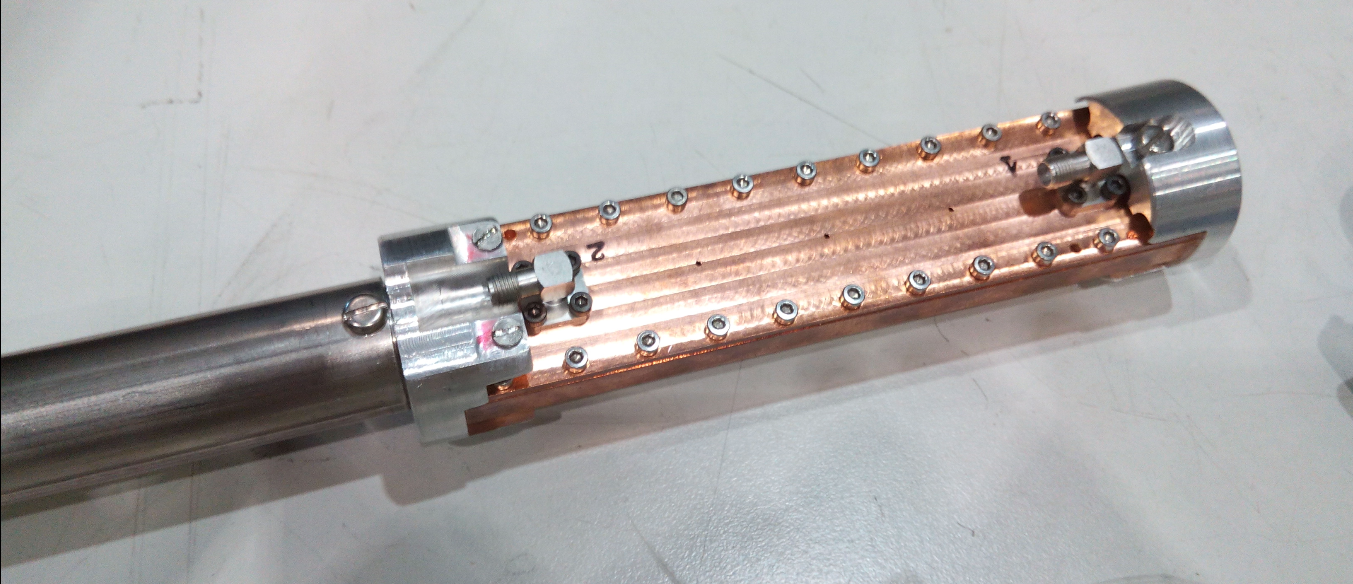}
\caption{Stainless steel {316L} fabricated prototype based on the five cavity X-band filter {design}.
Left: cavity before coating and assembly. Right: copper-coated cavity mounted onto the insertion stick. To place the cavity in the region of constant magnetic field in the CAST magnet,
it is attached onto a $\sim$2-m long hollow insertion stick through which the cabling can be guided to the flange,
cf. Fig~\ref{wsetup} (cables not shown in the picture).
}
\label{fig:cavityss}
\end{center}
\end{figure}

Due to the requirements imposed by the high magnetic field environment of CAST where the filter {is} placed, the designed filter has been manufactured using stainless steel 316L by a standard milling machining process, as can be seen in Figure~\ref{fig:cavityss}. A copper coating layer with approximately $30~\mu$m thickness has been applied to the structure
to improve the electrical conductivity.
This copper layer is expected to have a residual resistance ratio (RRR) between 30 and 200 {but at the frequencies of interest the anomalous skin-depth effect~\cite{Rogers:1988bq,2008IJIMW..29..924F} moderates the increase in conductivity, resulting in a lower increase in $Q$ than otherwise expected from the RRR.}
A higher quality factor could have been obtained by means of a silver outer layer, but this solution was ruled out since it would require a nickel layer between stainless steel and silver, which is incompatible with the high magnetic field environment of CAST. The effect of the stainless steel on losses is negligible since the thickness of the copper {layer} is much larger than the skin depth for copper at 8.4 GHz (0.7 $\mu$m at room temperature and 0.1 $\mu$m at 2~K).

A linear correction expansion coefficient has been applied in the manufacturing process to compensate the change in the physical dimensions due to the temperature expansion
from 2~K to room temperature. The expansion coefficients for the stainless steel can be found in~\cite{nist}.

We have added an output coaxial probe to extract RF power from the cavity, named port 1, and an input coaxial probe to inject a calibration signal for diagnostic purposes,
named port 2. {The probes are placed at the center of the top side of the first and last cavity.} Right angle 50 $\Omega$ SMA coaxial connectors \cite{coaxial} are used to fit in the CAST magnet borehole~\cite{Zioutas:1998cc}. The electromagnetic properties of the filter with probes have been computed with CST at 2~K and at room temperature. {The probe} in port 1 has been designed to operate at critical coupling, for which the tip has been adjusted to be level with the internal face of the cavity.  {The probe of port 2 is intended to be weakly coupled, and so its tip has been retracted 1~mm inside the internal wall of the cavity.}
The input coaxial probe will be short-circuited during axion detection operation.

\begin{figure}[h!]
\begin{center}
\includegraphics[width = 1\textwidth]{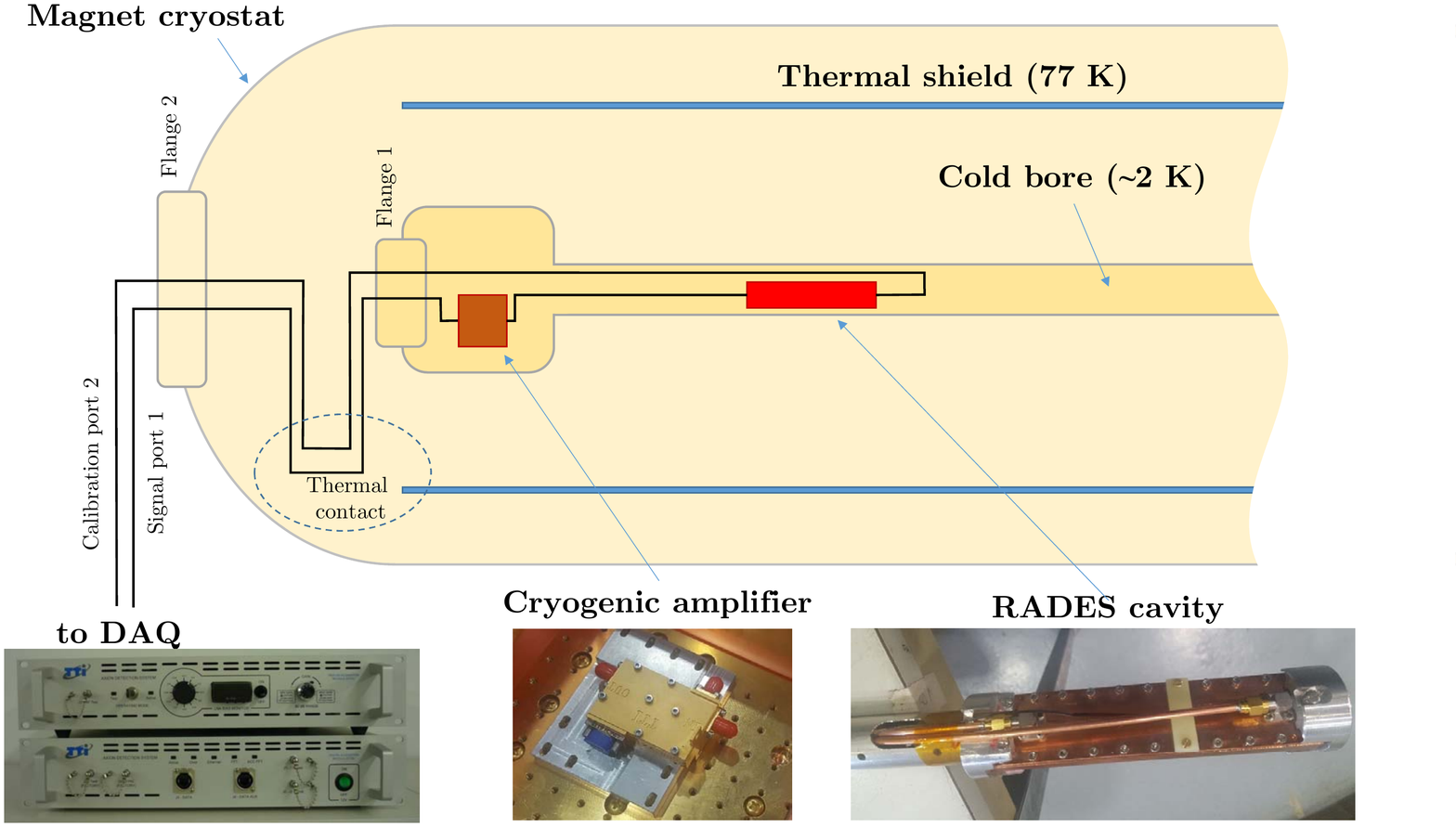}
\caption{Layout of the RADES setup inside the CAST magnet, showing the position of the cavity inside the magnet bore, the cryogenic amplifier and the transition of both RF connections (signal and calibration) from the cavity ports to the DAQ system outside the magnet.
}
\label{wsetup}
\end{center}
\end{figure}

The cavity is placed inside one of the bores of the prototype LHC dipole magnet placed at CAST at CERN. {Figure \ref{wsetup} shows the schematic layout of the setup inside the magnet. The signal is amplified at cryogenic stage and extracted to the DAQ electronics placed outside the magnet.} A cryogenic amplifier~\footnote{Model TXA4000 manufactured by TTI Norte \cite{ttinorte}} providing a 40~dB gain in the 8-9~GHz {range}, is placed inside a copper vessel at the cryogenic section limited by flange 1.
Then, RF cables are transitioned from cryogenic environment to room temperature by means of thermal plates. In addition, port 2 is intended for calibration and monitoring of the working frequency and correct operation of the amplifier.
Temperature and bias cables are made of phosphor bronze from \cite{lakeshore} to minimize thermal leakages. RF cables are 3.5 mm semirigid coaxial copper from Microcoax \cite{microcoax};
connectors are Sub-Miniature version A (SMA).

The Data Adquisition System (DAQ) is formed by the analog module (a heterodyne receiver) and the digital module (an A/D converter plus a field programmable gate array (FPGA)). The DAQ was manufactured by TTI Norte \cite{ttinorte}.  The analog module includes a low-noise amplifier operating from 8 GHz to 9 GHz, with a nominal gain of 55 dB and a very good return loss of 30 dB, and a submodule that makes the frequency translation from X-band to a intermediate frequency of 140 MHz.  The analog signal is converted into digital format  with a sampling rate of 37.5 MHz. The FPGA integrates 2048 Fast Fourier Transforms (FFTs) to store 12 MHz of bandwidth. {A more detailed description of the DAQ, is left for a future publication including our data analysis to search for an axion signal.}

\begin{figure}[hbtp]
\begin{center}
\includegraphics[height= 8cm]{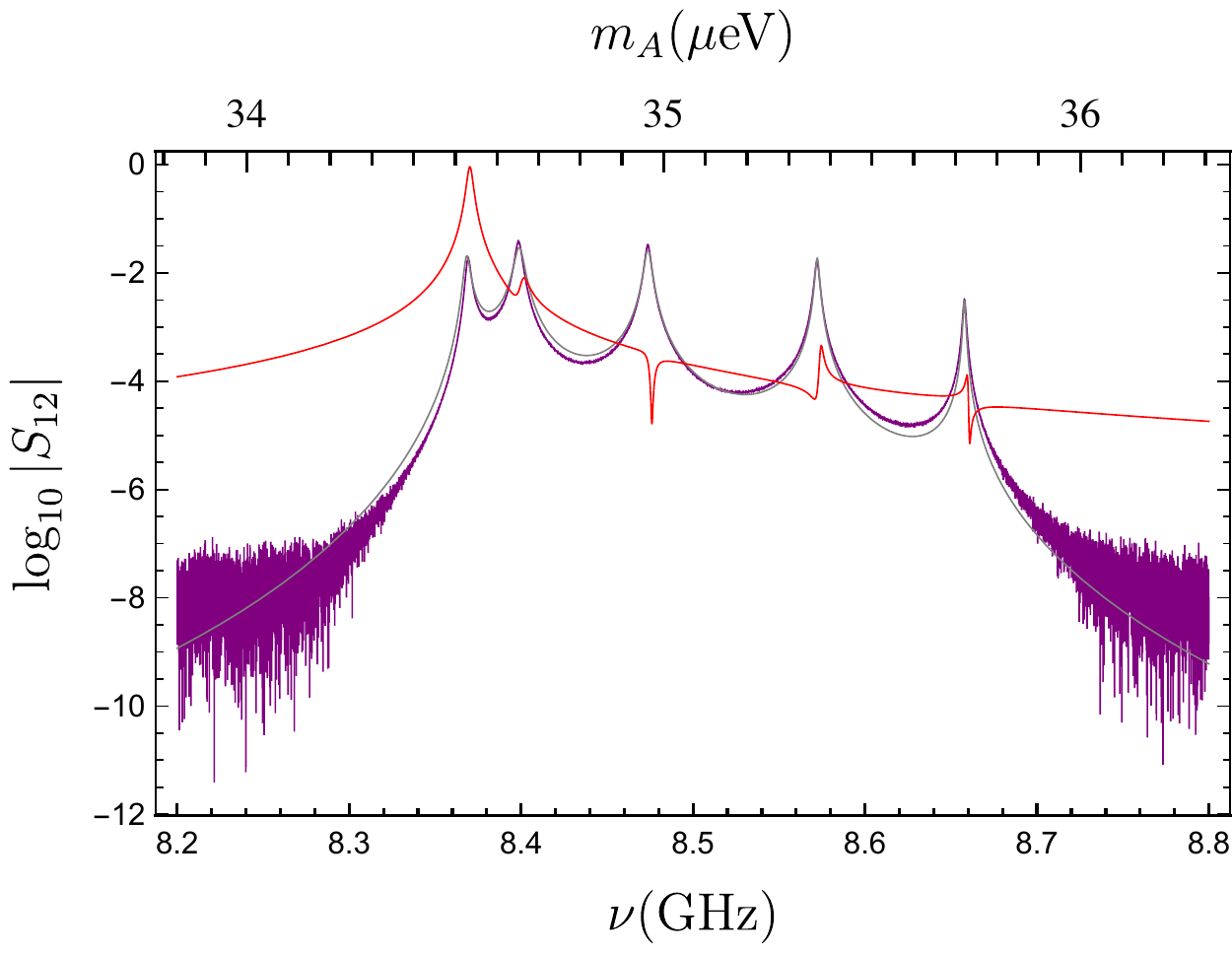}
\includegraphics[height= 8cm]{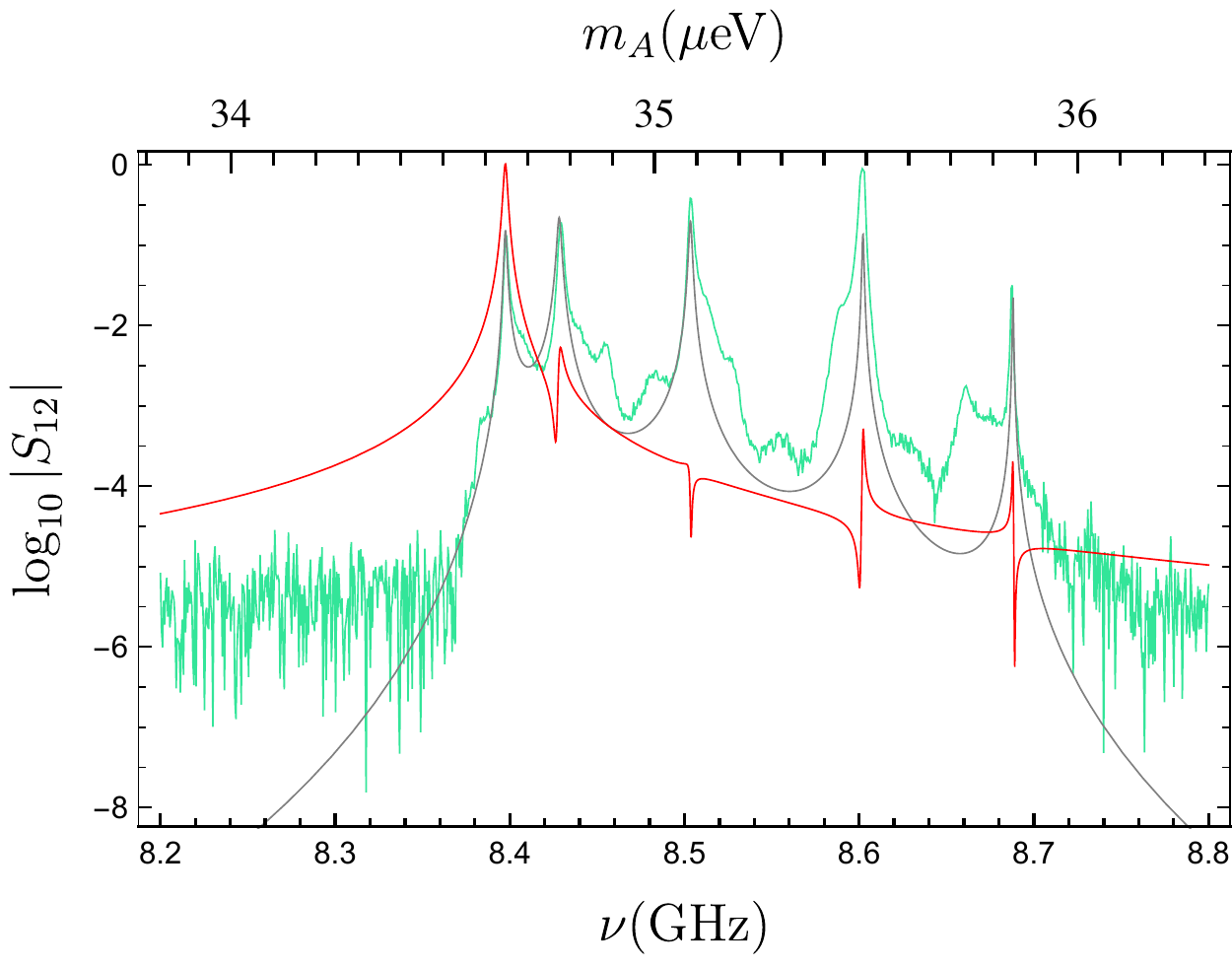}
\caption{Measured scattering transmission parameter $S_{12}$ (purple and green lines for room temperature and 2.13~K respectively), analytical model (gray lines) and axion DM power normalized to the peak (red). The RF signal has been injected through port 2 and detected in port 1. The upper plot shows room temperature results and the lower one at 2.13 K. In the latter case, the RF cables and amplifier are included in the system, which causes the extra spectral features other than the five resonant peaks.   }

\label{s12}
\end{center}
\end{figure}

{The cavity characterization data shown in this paper were taken with a} Vector Network Analyser (VNA) Rohde \& Schwarz ZVA67 (ZVA24) with built-in attenuators to input an X-band RF signal in port 1 at room temperature (298~K).
In Figure~\ref{s12} the measured transmission scattering parameter of the filter $S_{12}$ is shown at room temperature (upper plot) and at 2.13 K (bottom plot) for frequencies around 8.4 GHz. {In both cases the five resonant peaks of the filter are clearly seen. The slight frequency displacement of the peaks (see Table~\ref{tab:frequencies}) is compatible with the contraction of the dimensions at cryogenic temperatures. While the data at room temperature were taken at the laboratory with the cavity directly connected to the VNA, the data at 2.13 K were taken including the cables and amplifier and with the input signal properly attenuated at 0 dB to avoid saturation
of the cryogenic amplifier, 
which explains the additional spectral features in the plot.}
{The measured pattern of frequencies agree very well with the CST simulated ones, as seen in Table~\ref{tab:frequencies} and the absolute offset in frequency between data and simulation is within the mechanical tolerance of our fabrication. This gives us confidence that the actual field distribution of our cavity is indeed as predicted by the simulation. }

\begin{table}[hbtp]
\begin{center}
\begin{tabular}{|c|c|c|c|c|c|}\hline
Mode & $\fomega^{\rm 298~K}_i/(2\pi)$ & $\fomega^{\rm 2~K}_i/(2\pi)$  & $\fomega^{\rm CST}_i/(2\pi)$  & $(\fomega^{\rm 298~K}_i-\fomega^{\rm 2K}_i)/(2\pi)$ & $(\fomega^{\rm CST}_i-\fomega^{\rm 2K}_i)/(2\pi)$\\ \hline \hline
1		&	8.379	&	8.398	&	8.428	&	0.028	&	0.030	\\ \hline
2		&	8.399	&	8.429	&	8.454	&	0.031	&	0.025	\\ \hline
3		&	8.474	&	8.504	&	8.528	&	0.030	&	0.024	\\ \hline
4		&	8.572	&	8.602	&	8.625	&	0.030	&	0.023	\\ \hline
5		&	8.658	&	8.687	&	8.710	&	0.030	&	0.023	\\ \hline
\end{tabular}
\caption{{Resonant frequencies (in GHz) for the five modes of the cavity. Columns 1 and 2 show the experimental values obtained at 298~K and 2~K  respectively, while column 3 shows the values obtained from the CST simulation (from Table~\ref{tab:modes}). The difference between the 298~K and 2~K values (column 3) is a constant offset due to the temperature contraction. The difference between the experimental and simulated values is also approximately constant and compatible with the tolerance of our fabrication.  }}
\label{tab:frequencies}
\end{center}
\end{table}

{An additional check is done to validate our understanding of the system. The data of Fig.~\ref{s12} is also compared directly with the analytical model using Eq.~\ref{s12model}. All matrix parameters in~\ref{OMt} are allowed to vary independently to find the best fit (including an imaginary part for the diagonal elements to allow for cavity losses and properly fit the width of the peaks), as well as a overall normalization. So the system is allowed to depart from the optimal solution defined in section~\ref{sec:modelling}. The result of the fit to the 298~K data, shown as  gray line in Fig.~\ref{s12}, is able to reproduce the measured data remarkably. Then we check that the solution found in this way shows only a very mild departure from the optimal solution  and,
most importantly, its geometric factor is within $0.1\%$ of the maximum value expected for the optimal solution. The same solution is overlaid with the 2~K data, only readjusting the normalization and the $\fomega_1$ to account for the temperature contraction. For the current prototype, we consider that the cross-checks presented confirm our understanding of the system, and that the fabricated prototype enjoys a field distribution very similar to the one intended. Further work will go in the direction of better establishing this protocol of comparisons of experimental data with both simulations and analytical model, in order to assess the requirements in terms of mechanical tolerances required for larger, more demanding RADES prototypes. In addition, the output power due to axion DM can be computed immediately with the help of \eqref{genform}, obtaining the red curves of Fig.~\ref{s12}. We have normalized the curves to the maximum power on resonance.
As designed, only the first peak couples to the axion.   }

Using the shape of the 8.4 GHz peak, we have determined a loaded $Q$ value of $\sim 6000$,
which would imply $Q^u_{1}\sim 12000$ if the filter was critically coupled. From $|S_{11}|$ measurements in the lab we roughly estimate our actual port 1 coupling to be around 0.64,
which implies $Q^u_{1}\sim 16000$.
This is a factor of $\sim 3-4$ smaller than the predicted value shown in Tab.~\ref{tab:modes}, which we suspect
it is due to a smaller RRR ratio than assumed of our Cu coating due to the anomalous skin effect or other imperfections in the coating, and perhaps the effect of the horizontal cut in the cavity (not included in the simulations). Further work will be invested in controlling the output port coupling and modeling $Q$.

{To conclude, the behavior of the prototype satisfactorily matches our expectations from the analytical model and gives us confidence on the validity of the theoretical framework described in section~\ref{sec:modelling} to guide us in the design of more ambitious setups. The prototype described above is now in a few-weeks data taking phase in CAST. In a future publication we will report on the experience and results from this first data taking.} In the following section we anticipate the expected sensitivity of our setup to axions at a masses corresponding to around 8.4 GHz.

\section{Sensitivity projection \label{sec:sensi}}

In the following we give an estimate of
the prospect sensitivity of the RADES prototype cavity.
This estimate is based on geometric and electromagnetic properties
of the prototype cavity but
no data analysis of the data acquired with the cavity is pursued.
The analysis of the data taken with the RADES cavity will be the topic of a separate article.

The output power of the cavity when a mode $i$ is excited resonantly by axion DM, $m_A\sim \omega_i$, is given by \eqref{power1}
\bea
P &=& \kappa  \gagamma^2\frac{1}{m_A} B_e^2 \rho_{\rm DM} V Q_i { \geo^2_{i}}\\
  &=&
  1.25 \times 10^{-24}\, {\rm Watt} \frac{\kappa}{0.5} {C_{A \gamma}}^2
   \frac{m_A}{30\, \rm \mu eV} \left(\frac{B_e}{9\, \rm T}\right)^2 \frac{V}{1\, \rm l} \ \frac{Q}{10^4} \ \left(\frac{\geo_{i}}{0.69}\right)^2
   \label{signi}
\eea
where we have taken the local DM density $\rho_{\rm DM}= m_A^2A_0^2/2= 0.4$ GeV/cm$^3$. Using that the axion excitation
has a bandwidth $\Delta \nu_a \simeq m_A/(2\pi Q_a)$ with $Q_a \sim 10^6$, much smaller than the width of the cavity resonance $\Delta \nu_c \sim m_A/2\pi Q$. Here $\kappa$ the cavity coupling efficiency (see appendix) and ${\geo}_{i}$ is the geometric factor defined in \eqref{geo}, which for a filter becomes the sum \eqref{geot}. We have also used the QCD axion relation,
\begin{equation}
 \gagamma \equiv 2.0 \times 10^{-16} C_{A \gamma} \frac{m_A}{\mu {\rm eV}} {\rm GeV}^{-1} \ ,
\end{equation}
since we want to gauge our sensitivity through our reach on the dimensionless ${\cal O}(1)$ parameter $C_{A \gamma}$, cf.  \cite{Irastorza:2018dyq}.

The tiny axion signal needs to compete against the effective noise temperature of the system $T_{sys}$,
typically the sum of thermal and amplifier noise. In the axion line-width this corresponds to a power,
\begin{equation}
\label{nosi}
P_T = T_{sys} \Delta \nu_a 
= 6.0\times 10^{-19} \, {\rm Watt}  \frac{T_{sys}}{6\, \rm K}  \frac{10^6}{Q_a} \frac{m_A}{30\, \rm \mu eV} .
\end{equation}

The noise is expected to be smooth as a function of frequency and can be thus subtracted. The signal has to be only stronger than the expected noise fluctuations in the bin, which are $\sigma_{P_T}=P_T/\sqrt{\Delta\nu_a t}$ after a measurement time $t$.  Judging from \eqref{signi} and \eqref{nosi} we need circa $\Delta \nu_a t\sim 10^8$ to find an axion DM signal, which corresponds to measurement times of the order of  $t\sim $ few hours.
Demanding a meaningful signal to noise ratio $S/N$ for a given measurement time $t$ with the cavity tuned to a given axion mass, the sensitivity for the axion-photon coupling $C_{A \gamma}$ is then given by  Dicke's radiometer equation, $S/N=P_a  \sqrt{\Delta\nu_a t} /P_T$, as
\begin{eqnarray}
\left.C_{A \gamma}\right|_{\rm reach}
&\simeq & 21.7
\left(\frac{\frac{S}{N}}{3}\right)^\frac{1}{2}\frac{9\, \rm T}{B_e}\left(\frac{1\, \rm l}{V}\right)^\frac{1}{2}
\left(\frac{10^4}{Q}\right)^\frac{1}{2} \left(\frac{0.69}{{{\mathcal{G}_i}}}\right)^\frac{1}{2}
\left(\frac{T_{\rm eff}}{10\, \rm K}\right)^\frac{1}{2}
\left(\frac{0.5}{\kappa}\right)^\frac{1}{2}
\exclude{\frac{2.6\times 10^{-19}}{\theta}}
\left(\frac{m_A}{30\, \rm \mu eV}\frac{\rm hour}{t}\right)^\frac{1}{4} \nonumber .
\label{eq:sensi}
\end{eqnarray}

\begin{figure}[hbtp]
\begin{center}
\includegraphics[width = .5\textwidth]{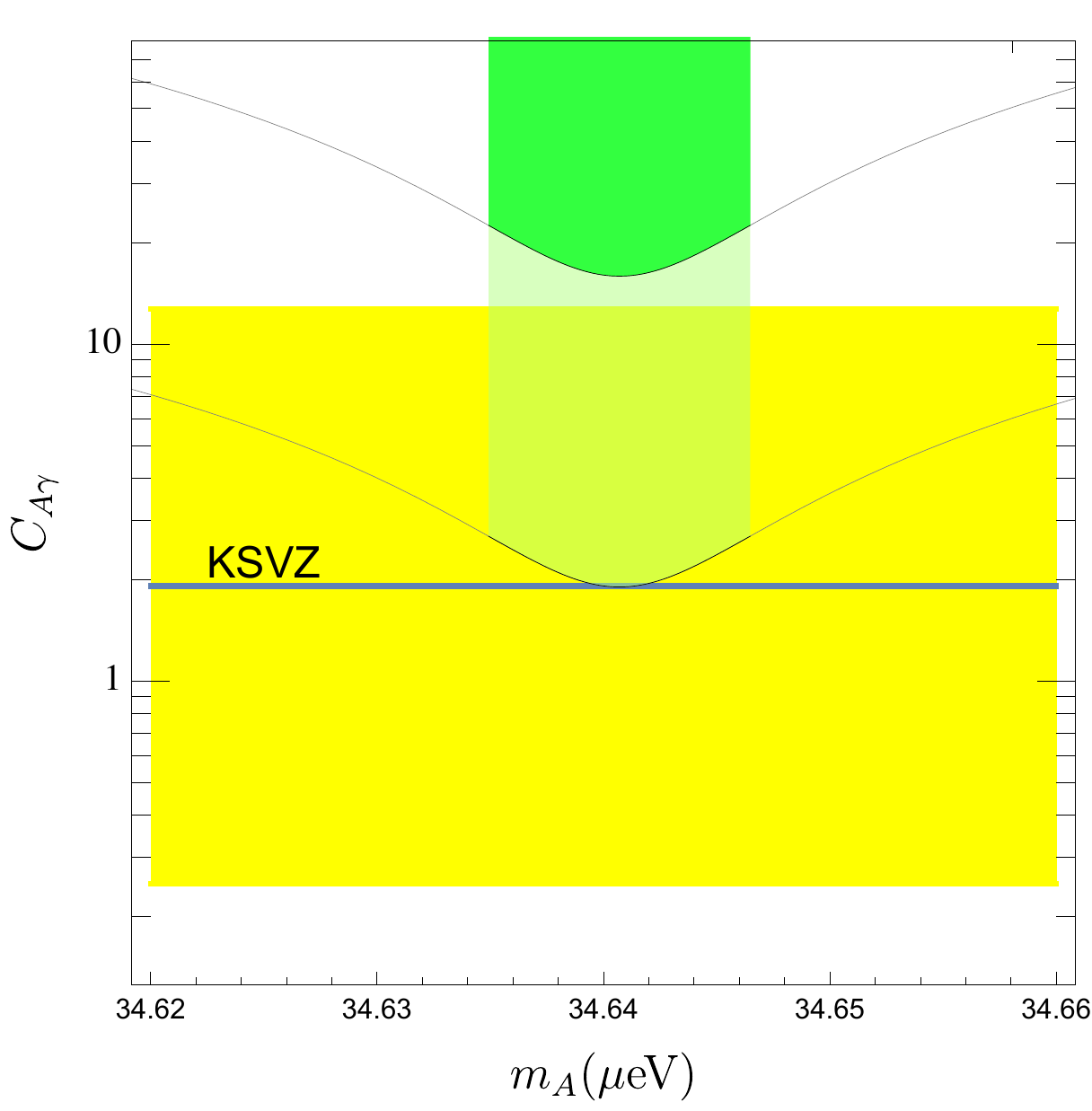}
\caption{Prospect sensitivity  (green region) to the axion photon coupling of the RADES prototype inside the CAST magnet
assuming 20 weeks data taking with electromagnetic properties of the filter detailed in this text.
Note that we cut the width of the green region at the half-width of the resonance peak.
A RADES-like filter of $\sim$350 sub-cavities filling a full CAST LHC magnet bore-length of 9~m would reach KSVZ sensitivity (light green region). }
\label{fig:prospect}
\end{center}
\end{figure}

The volume of the prototype cavity described in section \ref{sec:RADESdesign} is $V\simeq 0.03$ l.
As reasonable measurement time for the run of the prototype cavity we have assumed 20 weeks, the $Q$ value is taken to be 6000 and and effective noise temperature $T_{\rm eff}\sim$ 6~K (4~K from vendor test report and 2~K as magnet temperature).
At a signal to noise ratio of 3 we then obtain the prospect shown in Figure \ref{fig:prospect}.
It has to be emphasized that the prospect presumes that axions constitute all of Dark Matter.

Note that this sensitivity is obtained only for a very narrow axion mass range of order $\sim m_A/Q$.
A resonant mode with a given $Q$ has a line width which contains therefore a number $\sim Q_a/Q$ of possible axion mass channels, so that with the cavity tuned to a given frequency we are measuring all these channels simultaneously.
For a future tunable RADES cavity, tuning steps are then of order $\nu/Q$.


In Figure \ref{fig:prospect}, we have also included
the benchmark sensitivity of KSVZ axions $|C_{A \gamma}|=1.92$
and a yellow band containing QCD axion models as summarised in~\cite{DiLuzio:2016sbl, Irastorza:2018dyq}.
In this estimate the sensitivity of our prototype is already at the level of the most optimistically coupled models in the band, and within a factor of a few in $\gagamma$ to the KSVZ
theoretical prediction~\cite{Shifman:1979if,Dine:1981rt}, assuming the electromagnetic properties
already measured at 2.13~K. The results of the cavity performance presented in this work are encouraging us to build a larger cavity which can reach benchmark QCD sensitivity.

\section{Summary and Conclusions \label{sec:summary}}


{There is a strong motivation to search for axion DM in the}
10-100 $\mu$eV mass range.
{An increasing experimental effort is taking place worldwide to develop competitive implementations of the axion haloscope technique in this mass range, for which the $V \sim m_A^{-3}$ relationship of a simple resonant cavity leads to a strong penalty in sensitivity. Some of these efforts focus on developing extended resonant structures that could instrument large magnetic volumes $V$ while resonating at relatively high frequencies~\cite{TheMADMAXWorkingGroup:2016hpc,Rybka:2014cya,Miceli:2015xas}. The RADES approach developed in this paper} is based on the geometry of microwave filters to achieve this goal.

We have presented the theoretical framework to characterize the resonant modes in an array of long-rectangular cavities
segmented and connected through irises. {The framework has similarities with the cavity array developed in~\cite{Goryachev:2017wpw}, although it differs in the formalism used. It allows to build an analytical model that provides practical design guidelines} to find the optimal cavity parameters maximizing the $\mathcal{G}^2$ factor. {We have designed and built a concrete implementation of the concept, as a 5-cavity filter-like structure, that enjoys an optimized resonant mode at 8.4 GHz}. We have ran numerical simulations based on the Finite Integration Technique, to assess and fine-tune the final parameters of the filter-like structure.

{A first RADES prototype, following the above prescriptions, has been} built in stainless steel 316L with a copper coating layer of $\sim30~\mu$m thickness.
We have measured the electromagnetic properties of the cavity both at room temperature and at 2.13~K inside the CAST magnet at CERN. The observed parameters agree very well both with simulations and with the simple analytical model, validating the method presented in this work. {This first RADES prototype is now installed inside the CAST magnet and actual data-taking with the magnet powered is ongoing with a dedicated DAQ system. Preliminary sensitivity prospects have been presented. Even with the small volume of the current prototype, sensitivity to the optimistic edge of the QCD axion band should be already achievable (for a thin range centered around $m_A \sim 34.64 \mu$eV), while sensitivity to KSVZ would be reachable by a larger version filling all the 10 m length of the CAST magnet.}

The results here presented demonstrate the potential of microwave filters based on coupled adjacent cavities as axion haloscopes from C-band to K-band frequencies. {Our next steps are to design} larger $V$ filters which can cover the QCD motivated values of {$\gagamma$, as well as to devise a suitable tuning mechanism to allow for effectively scan a relevant $m_A$ range.}

\section*{Acknowledgements}
{
This work has been funded by the Spanish Agencia Estatal de Investigaci\'on (AEI) and Fondo Europeo de Desarrollo Regional (FEDER) under project FPA-2016-76978, and Fundacion Seneca Ref. 20147/EE/17, and was supported by the CERN Doctoral Studentship programme. IGI acknowledges also support from the European Research Council (ERC) under grant ERC-2009-StG-240054 (T-REX project). JR is supported by the Ramon y Cajal Fellowship 2012-10597, the grant FPA2015-65745-P (MINECO/FEDER), the EU through the ITN ``Elusives'' H2020-MSCA-ITN-2015/674896 and the Deutsche Forschungsgemeinschaft under grant SFB-1258 as a Mercator Fellow. CPG was supported by PROMETEO II/2014/050 of Generalitat Valenciana, FPA2014-57816-P of MINECO and by the European Union's Horizon 2020 research and innovation program under the Marie Sklodowska-Curie grant agreements 690575 and 674896.

We want to thank Dr. M. Guglielmi and Prof. V. E. Boria from Universidad Polit\'ecnica de Valencia for early discussions that led to the consideration of microwave filters as axion resonant cavities. We wish also to thank our colleagues at CAST and CERN for their support and advice in specific aspects of the project, most especially to W. Funk, to J.M. Laurent and the surface treatments workshop for their help with the Cu coating, to X. Pons and the cryolab team for their help with the RF cables and their cryogenic transition, to F. Caspers, for his many advices, and to L. Miceli for sharing his experience in CAST-CAPP, the other haloscope project in CAST. We also thank A. Sulimov from DESY for helpful discussions on the field distribution inside the cavity. We finally thank Jose M. Catal\'a-Civera from ITACA, Universitat Polit\`ecnica de Val\`encia for his advices in the manufacturing of the filter.
}

\appendix

\section{General formalism} \label{sec:genform}
In the background of a time-varying axion DM field\footnote{The spatial variation of the axion DM field is negligible as long as the number of cavities is $N\lesssim 1000$. Much longer cavities could be used to infer the velocity distribution of axion DM as pointed out in~\cite{Irastorza:2012jq}.}, $A(t)$, and a strong magnetic field, ${\bf B}_{e}$, Maxwell's equations get an additional source
\bea
\nabla\cdot {\bf E} =  0
\quad &, & \quad
\nabla\times {\bf B}-\dot{\bf E} =  \gagamma {\bf B}_{e}\, \dot A\\
\nabla \cdot {\bf B}  = 0
\quad &, & \quad
\nabla\times {\bf E}+\dot{\bf B}  = 0
\eea
due to the axion coupling to two photons, which is described by the Lagrangian density,
\be
{\cal L}_{A\gamma} = \gagamma
{\bf B}\cdot {\bf E} \, A \, .
\ee
Let us first review how this source excites a resonant cavity.
The $\bf E, {\bf B}$ fields can be expanded as a sum of orthonormal cavity modes $\Emod_m({\bf x})$ that solve the Poisson equation, $\nabla^2 \Emod_m({\bf x}) = -\omega_m^2 \Emod_m({\bf x})$ with a characteristic  eigenfrequency $\omega_m$. Modes are normalised as $\int_V d^3{\bf x}\, \Emod_m \cdot \Emod_m'=V \delta_{mm'}$ where $V$ is the volume of the cavity.
Writing the electric field as ${\bf E}=\sum_m E_m(t) \Emod_m({\bf x})$, Ampere's equation projected into the $m$-th mode gives the time evolution of the amplitude
\be
\ddot E_m + \omega_m^2E_m +\Gamma_m \dot E_m=  - \gagamma B_e \ddot A\,  \cal G \,
\ee
where we have parametrised energy losses by a decay rate, $\Gamma_m$, and defined the geometric factor,
\be
\label{geo}
\geo_m = \frac{1}{B_e V} \int_{V_c} d^3{\bf x}\,   {\bf B}_e \cdot \Emod_m \, .
\ee
Observe that the background ${\bf B}_e$ field must have a parallel component along the desired mode's $E$-field to become excited by axion DM.  We will use homogeneous  ${\bf B}_e$ fields, so it is convenient to use $B_e = |{\bf B}_e|$. The decay rate is defined such that, in absence of sources and for small losses $\Gamma_m\ll \omega_m$ the field amplitude decrease as  $|E_m(t)| = |E_m(0)|\exp(-\Gamma_m t/2)$ and the energy in a mode $U_m$ as $|E_m|^2\propto \exp(-\Gamma_m t)$. The ratio
$\omega_m/\Gamma_m$ corresponds to the energy loss per oscillation cycle of the mode $m$ and is defined as the quality factor of a resonator,
\be
Q_m=\frac{\omega_m}{\Gamma_m}.
\ee

When excited by a monochromatic axion DM field, $A=A_0 e^{-\ii \omega t}$,
the $E$-field amplitude of each mode approaches the steady state solution,
\be
\label{Em1cav}
E_m= - \gagamma B_e A  \times \frac{ \omega^2 \geo_m  }{\omega^2-\omega_m^2 + \ii\omega \Gamma_m}  .
\ee
A quick look at Ampere's equation, suggests that the typical $E$-field amplitude induced by the axion DM field is $E\sim \gagamma B_e A$.
This is exactly what we get when we excite a mode much above its natural frequency, $\omega\gg \omega_m$ (barring the geometric factor). Below the natural frequency, $\omega<\omega_m$, $E_m$ gets suppressed by a factor $\omega^2/\omega_m^2$. On resonance $\omega\sim \omega_m$ the amplitude increases by a factor $(\omega_m/\Gamma_m)^2=Q_m^2$.
The EM energy stored in the cavity splits in a sum over modes,
\be
U = \int d^3{\bf x} \frac{1}{2}(|{\bf E}|^2+|{\bf B}|^2) = \sum_m \frac{1}{2}|E_m|^2 \(\frac{\omega^2+\omega_m^2}{2 \omega^2}\) = \sum_m U_m .
\ee
The energy in a mode can be read by a suitably coupled small antenna, but the power extracted contributes to the losses, i.e. $\Gamma_m=\Gamma^c_m + \Gamma^s_m$ where $\Gamma^c_m$ represents damping due to surface currents in the cavity walls or other intrinsic loses and $\Gamma^s_m$ the losses invested in the output signal. The signal power is,
\be
\label{power1}
P = \Gamma^s_m U_m =
\kappa \frac{\omega_m}{Q_m}   \frac{ |\gagamma B_e A_0|^2 \, V|\geo_m|^2 }{2}\,
\frac{\omega^4 }{(\omega^2 -w^2_m)^2  +(\omega_m \omega/Q_m)^2 }  ,
\ee
where the coupling coefficient is $\kappa = \Gamma^s_m/(\Gamma^c_m + \Gamma^s_m)$. For a given value of the intrinsic losses, the optimum signal is obtained for $\kappa=1/2$. On resonance the output power gets enhanced by the quality factor $Q_m$.

Let us now consider a number $N$ of cavities. In this paper we focus on the case where all the cavities have one mode close to a common central resonant frequency, which is well separated from neighbouring resonances and couples to the axion DM with geometric factors of order $\geo_c\sim 1$.
From this moment on, neighbouring modes are integrated out of the discussion, assuming they play no role. Each cavity has thus just one mode.
To make our notation more compact, we label the amplitude of the relevant mode, $r$, of the $q$-th cavity as $\E_q$, and introduce complex frequencies as
\be
\qomegaC^2_q = \omega_{q}^2-\ii \omega_{q} \Gamma_{q},
\ee
including the losses in the imaginary part and using $\omega\sim \omega_r$ there for practicality.
We couple the cavities through small irises forming a linear array that we call filter. The coupling is linear and can be described with a coupling coefficient $\kk_{qq'}$.
When excited by a monochromatic axion DM field $A=A_0 e^{-\ii\omega t}$,
the system of coupled equations for the amplitudes of the fundamental mode can be described by
\begin{align}
\nonumber
(\omega^2 -\qomegaC_1^2)\E_1 &= \kk_{12}\E_2   -  \gagamma B_e A\, \omega^2\,  \geo_1 \\ \nonumber
(\omega^2 -\qomegaC_2^2)\E_2 &= \kk_{21}\E_1 + \kk_{23}\E_3  - \gagamma B_e  A\, \omega^2\,  \geo_2 \\ \nonumber
&... &\\
(\omega^2 -\qomegaC_N^2)\E_N &= \kk_{N,N-1}\E_{N-1}   - \gagamma B_e A\, \omega^2 \,  \geo_N ,
\end{align}
which we can write as the vector equation \eqref{theequationt},
\be
\label{theequation}
(\omega^2 {\mathbb{1}} - \MM)\cev{\E} = \cev {\jA} = - \gagamma B_e A_0\,  \omega^2 \, \cev \geo .
\ee
Note that we use overbars for vectors of cavity properties, and boldface for 3D vectors like electric or magnetic fields.

An array of rectangular cavities segmented and connected through irises is modelled by the
tri-diagonal matrix
\begin{equation}
\label{OM}
\MM =
\left(
\begin{array}{c c c c c c}
\qomegaC_{1}^{2} & \kk_{12} & 0 & 0 & 0 & 0 \\
\kk_{21} & \qomegaC_{2}^{2} & \kk_{23} & 0 & 0 & 0\\
0 & \kk_{32} & \qomegaC_{3}^{2} & \kk_{34} & 0 & 0 \\
0 & 0 & \ddots & \ddots & \ddots & 0 \\
0 & 0 & 0 & \ddots & \ddots & \ddots \\
0 & 0 & 0 & 0 & \kk_{N,N-1} & \qomegaC_{N}^{2} \\
\end{array}
\right)  ,
\end{equation}
In practice we will mostly consider real $\kk$'s with $\kk_{qq'}=\kk_{q'q}$, neglecting losses.
\noindent
The EM modes of the filter around the fundamental mode correspond to the $N$ eigenvectors of this matrix,
$\{\cev \eig\}$,  satisfying
\be
\label{eigenformula}
\MM\, \cev \eig_i = \lambda_i \, \cev \eig_i .
\ee
As long as imaginary parts are small, the matrix $\MM$ is symmetric and the eigenvectors approximately orthogonal in the ordinary sense.
Likewise, the $i$-th eigenvalue $\lambda_i$ correspond to the square of the $i$-th characteristic resonant frequency of the whole set of $N$-coupled cavities, which we label as $\lambda_i = \fomegaC^2_i$.
We will use the subindex $q$ to label properties of the individual cavities and $i$ for the global solutions of the filter array.  %

As a first and very illustrative example, we consider an array of identical cavities coupled by identical irises.
The $\Omega$ matrix is Toeplitz with identical diagonal elements $\qomegaC_q^2$ and couplings $\kk$, and can be immediately diagonalised.
The eigenvectors and frequencies are
\be
\cev \eig_i = \frac{1}{\sqrt{(1+N)/2}}
\vvvv{\sin\(\frac{i \pi}{N+1}1\)}{\sin\(\frac{i \pi}{N+1}2\)}{...}{\sin\(\frac{i \pi}{N+1}N\)}
\quad , \quad
\fomegaC^2_i = \qomegaC^2_q + 2 \kk \cos\(\frac{i\pi}{N+1}\) \quad ; \quad i=1,...,N .
\ee
We note that this solution is also shown in \cite{Goryachev:2017wpw}, barring differences in formalism. The result is valid for arbitrary complex $\qomegaC_q^2$ and real $k$ (although it generalises straightforwardly to the complex case).
The original resonance splits into $N$ non-equally spaced modes in a band $\Delta \omega  \simeq \kk/\omega_q$ centred at $\omega_q$.
For $k<0$, 
the lowest frequency mode corresponds to $i=1$ and its eigenvector has all positive components, i.e. the electric fields of the cavities oscillate in phase. As $i$ and the eigenfrequency increase, the E-fields alternate between positive and negative signed values faster until the $N$-th mode, for which the $E$-field changes sign in each contiguous cavity.
Therefore, we expect that the fundamental mode is the one coupling best to axion DM in an homogeneous $B_e$ covering all the array. We have normalised the eigenvectors as $|\cev e_i|^2=1$.
Note also that the normalisation factor decreases as $i$ increases.

\vspace{0.5cm}
Let us come back to the general case \eqref{theequation}. We can solve for the electric fields in each individual cavity
excited by axion DM \eqref{theequation} with the aid of the eigenmodes of the cavity.
Denoting as $R$ the transformation that diagonalises $\MM$
\be
R^{-1}\MM\, R =\OO^2 \equiv {\rm diag}\{\fomegaC_1^2,\fomegaC_2^2,...,\fomegaC_N^2\}
\ee
we formally obtain
\bea
\cev \E = 
R\(\frac{1}{\omega^2-\OO^2}\) R^{-1}\,  \cev {\jA}  .
\eea
\exclude{By definition, the $i$-th column of $R$ is the eigenvector $\vec x_i$ so $R_{ij}=(x_i)_j$
(apply vectors of one 1 to $\Omega R = \Lambda R$)}
which can be written as
\bea
(\cev \E)_q
		&\simeq& \sum_{i} (\cev \eig_i)_q\(\frac{\cev \eig_i \cdot \cev {\jA}}{\omega^2-\fomegaC_i^2}\)
\eea
where we have used that $R_{q i}=(\cev \eig_i)_q$, i.e. the transformation matrix is a row of column eigenvectors $\cev \eig_i$.
The approximation $\simeq$ is due to the fact that we have also used $R^{-1}\simeq R^T$.
Both matrices are exactly the same when $\Omega$ is real, and approximately equal when the imaginary components (due to losses being very small) are small.

The interpretation of the above formula is clear when we compare it with its 1-cavity counterpart, \eqref{Em1cav}. The original fundamental mode has split into $N$ modes of the array and so the electric-field in the $q$-th cavity, $(\cev \E)_q$, is now a linear superposition of the electric-fields of each array mode $i$ in that cavity. Each array mode contribution is weighted by the resonator response factor $(\omega^2-\fomegaC_i^2)^{-1}$ and the corresponding geometric factor $\cev  \eig_i \cdot \cev \geo$. This last quantity is precisely what one would expect from a geometric factor because the sum of individual integrals can be understood as a global $\bf \Emod \cdot {\bf B}_e$ integral.
In other words, the axion DM field oscillating at $\omega$, excites every mode of the array with a weight given by
$\cev  \eig_i \cdot \cev {\jA}/(\omega^2-\fomegaC_i^2)$. The electric field in a cavity of the array is the superposition of the
$E$-fields of each mode.
Note that all modes oscillate at the same frequency, being forced by the axion field, but they can have different phases and can cancel partially or totally. The interference is dictated by the sign of the geometric factor and the sign of $\omega^2-\fomegaC_i^2$ (whether $\omega$ is above the resonant frequency $\omega_i$ or below).

If the modes of the filter are sufficiently separated, the signal power output from a given port at a frequency where one mode dominates is still given by \eqref{power1} but $\kappa, \omega_m, Q,V, \geo$ have to be understood as pertaining a mode of the filter. Let us discuss the relation of the filter properties with the individual properties of each cavity.
The eigenfrequencies $\Omega_i^2$ are given by the diagonalisation of the matrix $\MM$ and so are the imaginary parts.
However, if the imaginary parts are very small, as will be our case, one can diagonalise Re$\{\MM\}$ with an orthonormal basis $\{\cev \eig\}$ and compute the losses of the $i$-th mode as a perturbation
\bea
\Gamma_ i \simeq (\cev \eig_i)^T{\rm Im}\{\MM\}\cev \eig_i
=  \sum_q \Gamma_q (\cev \eig_i)_q^2 ,
\eea
which follows from the definition \eqref{eigenformula} because Im$\{\MM\}={\rm diag}\{\omega \Gamma_1,...,\omega \Gamma_N\}$ is a diagonal matrix.
The last formula has the obvious interpretation of a reweighed sum of losses according to the energy stored in each cavity.
If the cavities are very similar, as will be our case, the intrinsic losses in each of them are very similar,  $\Gamma^c_q\sim \Gamma^c_0$,  and thus we obtain
\be
\Gamma^c_i \sim \Gamma^c_0 \sum_q  (\cev \eig_i)_q^2 = \Gamma^c_0 \, .
\ee
Therefore the \emph{unloaded} quality factor of the filter, $Q^u_i = \omega_i/\Gamma^c_i$, must be very similar to that of each of the individual cavities. The optimal coupling factor continues to be $\kappa = 1/2$, which means $\Gamma^s_i=\Gamma^c_i$.
However, the intrinsic losses are shared among the cavities and the output port will be placed in one of them, at least that would be the simplest option. It is good to bear in mind that this means that this output cavity could have a much larger ratio of output to intrinsic losses, which could potentially lead to large mode distortions and eventual quenching in the large $N$ limit.
The geometric factor for one cavity is defined in \eqref{geo}, which in the multi-cavity case can be interpreted as
\be
\label{geoN}
\geo_i =
\frac{\sum_q  V_q\,  \cev \eig_i \cdot \cev \geo}{V} ,
\ee
where $V=\sum_q V_q$ is the sum over cavity volumes. In the case where cavities are similar and their individual geometric factors too, $\geo_i\simeq \geo_0$.
The obvious boost in signal comes essentially from the increase in volume $V= \sum_q V_q$ which in the filter case will be $NV_q$.
In \eqref{geoN}, we have assumed an homogeneous ${\bf B}_e$ but it is straightforward to include if it varies between cavities.

If the modes of the cavity are not well separated or we want to take interference effects into account, we can derive a more general formula from the power output from a port in the $q$-th cavity,
\be
\label{genform}
P = \Gamma^s_q   \frac{ |\gagamma B_e A_0|^2 \, V_q}{2}
\left| \sum_{i} (\cev \eig_i)_q\(\frac{\cev \eig_i \cdot \cev {\jA}}{\omega^2-\fomegaC_i^2}\) \right|^2 .
\ee
This is the equation we use in our comparisons with experiment.

This equation is also valid for the power output when we artificially inject a signal in one or several cavities.
The $S_{12}$ parameter is defined as the power transmitted across a filter and can be used to calibrate our filters.
In order to compute it, we inject a signal in one cavity ($q_{\rm in}$ typically $q=N$ for us) and detect it in other (typically $q_{\rm out}=1$), the port we use for the axion DM signal.
This can be modeled by a source term in \eqref{theequation} that is not homogeneous like the axion, but localised in the input cavity
$\cev {\jA} \to \cev {J_{in}} $ with $(\cev {J_{in}})_q \propto \delta_{q,q_{\rm in}}$
The standard nomenclature is to call these ports 1 and 2.
Mixing a bit the notation to please logic and tradition simultaneously we can write,
\be
\label{s12model}
\left| S_{12} \right|^2=
\propto  \left| \sum_{i} (\cev \eig_i)_{q_{\rm out}}\(\frac{\cev \eig_i \cdot  {\cev {j_{in}}}}{\omega^2-\fomegaC_i^2}\) \right|^2 \propto
\left| \sum_{i} \frac{(\cev \eig_i)_{q=1} (\cev \eig_i)_{q=2}}{\omega^2-\fomegaC_i^2} \right|^2 .
\ee
The proportionality factor includes the strength of the input coupling, etc.

\bibliographystyle{utphys}
\bibliography{axionDM,additional}

\end{document}